EUROPEAN ORGANIZATION FOR NUCLEAR RESEARCH

# MEASUREMENT OF THE POLARIZATION OF $\Lambda^0$, $\overline{\Lambda}^0$, $\Sigma^+$ AND $\Xi^-$ PRODUCED IN A $\Sigma^-$ BEAM OF 330 GeV/c

The WA89 Collaboration

M.I. Adamovich[7], E. Albertson[4], Yu.A. Alexandrov[7], D. Barberis[2], C. Bérat[3], W. Beusch[1], K.H. Brenzinger[4,a], S. Brons[4], W. Brückner[4], A. Brunengo[2], M. Buénerd[3], F. Charignon[3], J. Chauvin[3], E.A. Chudakov[6], M. Dameri[2], F. Dropmann[4], J. Engelfried[5,b], F. Faller[5,l], B.R. French[1], S.G. Gerassimov[4], M. Godbersen[4], P. Grafström[1], J.Y. Hostachy[3], R.B. Hurst[2], Th. Kallakowsky[4,d], S. Kluth[5,k], H.M. Lauber[4,m], P. Lennert[5], S. Ljungfelt[5,e], K. Martens[5], Ph. Martin[3], R. Michaels[4,c], U. Müller[6], C. Newsom[f], B. Osculati[2], S. Paul[4], B. Povh[4], M. Rey-Campagnolle[3], H. Rieseberg[5], K. Röhrich[4,j], G. Rosner[6], L. Rossi[2], H. Rudolph[6,g], C. Scheel[h], L. Schmitt[4], H.-W. Siebert[5], A. Simon[5], A. Trombini[4], R. Touillon[3], B. Volkemer[6], Th. Walcher[6], G. Wälder[5], A. Wenzel[4,i], M.V. Zavertyaev[7]

[1]*CERN; CH-1211 Genève 23, Switzerland.*
[2]*Genoa Univ./INFN; I-16146 Genova, Italy.*
[3]*Grenoble ISN; F-38026 Grenoble, France.*
[4]*Heidelberg Max-Planck-Inst. für Kernphysik#; D-69117 Heidelberg, Germany.*
[5]*Heidelberg Univ., Physikal. Inst.#; D-69120 Heidelberg, Germany.*
[6]*Mainz Univ., Inst. für Kernphysik#; D-55099 Mainz, Germany.*
[7]*Moscow Lebedev Physics Inst.; RU-117924, Moscow, Russia.*

## Abstract

The polarization of $\Lambda^0$ , $\overline{\Lambda}^0$ , $\Sigma^+$ and $\Xi^-$ inclusively produced in $\Sigma^-$ induced interactions at 330 GeV has been measured in the experiment WA89 at CERN. This is the first measurement of polarization of baryons produced by a hyperon beam. No polarization of $\overline{\Lambda}^0$ is observed, as was also the case in proton beam data. At transverse momenta of about 1 GeV/c $\Lambda^0$ and $\Sigma^+$ show little polarization, significantly lower than in the proton beam data, while $\Xi^-$ have a polarization comparable to the polarization of $\Lambda^0$ produced in proton beams.

---

#) supported by the Bundesministerium für Forschung und Technologie, Germany, under contract numbers 05 5HD15I, 06 HD524I and 06 MZ5265

[a] Now at Inst. für Kernphysik; Mainz Univ.,D-55099 Mainz ,Germany.
[b] Now at FNAL, PO Box 500 Batavia, IL 60510, USA.
[c] Now at CEBAF, 12000 Jefferson Ave., Newport News, VA 23606, USA.
[d] Now at Siemens AG, Geschäftsbereich ÖN, D-81359 München, Germany.
[e] Now at PSI, CH-5232 Villigen, Switzerland.
[f] University of Iowa, Iowa City, IA 52242, USA.
[g] Now at LBL, MS 50D, Berkeley, CA 94720, USA.
[h] NIKHEF; PO Box 41882, NL-1009 DB Amsterdam, The Netherlands.
[i] Now at Université de Genève, CH-1228 Genève, Switzerland.
[j] Now at Inst. für Kernphysik, D-52425 Jülich, Germany.
[k] Now at Cavendish Lab., University of Cambridge, CB3 0HE Cambridge, UK.
[l] Now at Fraunhofer Inst. für Solare Energiesysteme, D-79100 Freiburg, Germany.
[m] Now at Heidelberg Univ.; Physikal. Inst., D-69120 Heidelberg, Germany.

# 1 Introduction

Since the discovery of a considerable polarization of $\Lambda^0$ hyperons produced inclusively by high energy protons [1] the phenomenon of hyperon polarization has been extensively studied, predominantly in proton beams at various energies.

It has been observed that nearly all hyperons inclusively produced by proton beams are polarized at high transverse momenta $p_T$. The absolute value of the polarization grows with $p_T$ in the region $0-1$ GeV/c and reaches values of about $-0.20$ for $\Lambda^0$ [1] , $+0.15$ for $\Sigma^-$ and $\Sigma^+$ [2, 3, 4] and $-0.10$ for $\Xi^-$ [5]. Generally the polarization also grows with the Feynman variable $x_F$. No significant dependence of the $\Lambda^0$ polarization on the beam momentum is observed from 12 GeV/c up to 2000 GeV/c (the equivalent momentum for ISR interactions) [6, 1, 7], while the $\Xi^-$ polarization is found to be slightly larger at $p_{beam}$ = 800 GeV/c than at $p_{beam}$ = 400 GeV/c [5, 8]. Measurable polarization has been observed for some anti-hyperons, such as $\overline{\Xi}^+$ [10] and $\overline{\Sigma}^-$ [4], but not for $\overline{\Lambda}^0$ [9].

Some measurements have been made with different beam particles, but this area has not been well investigated. The polarization of $\Lambda^0$s produced in a neutron beam has been found to have the same features as in proton beams [11]. In pion beams the polarization of $\Lambda^0$s produced in the forward hemisphere has a low measured value [12, 13, 14]. The existing measurements suggest that the polarization of $\Lambda^0$s is associated with the production of an s-quark from the sea and its attachment to the spectator quark or diquark from the projectile. In order to study the role of different quark flavours or quark masses in the polarization mechanism it is of special interest to measure the polarization of hyperons produced by strange particle beams. $\Lambda^0$ polarization has been measured in $K^-$ beams at energies from 4.2 to 176 GeV [15, 16]. A value of about $+0.50$ was observed in the forward hemisphere at $x_F > 0.3$. In this kinematic region one expects the s-quark from the projectile to be a spectator in the $\Lambda^0$ production process, the additional quarks being produced from the sea.

Hyperons produced exclusively in reactions close to diffraction dissociation are polarized even more strongly than in inclusive reactions. This had been observed at beam energies of several GeV [17], before the polarization of inclusively produced hyperons was found, and holds up to the ISR energies [18].

Perturbative QCD does not explain these polarization phenomena [19]; however, it is not considered to be applicable in the kinematic range of $p_T < 2$ GeV/c. A number of phenomenological models have been suggested using different methods to treat fragmentation processes and taking into account the spin wave functions of particles. These models give qualitative and semi-quantitative agreement with the observed polarizations. We mention here the models which give explicit predictions for hyperon beams. The semi-classical Lund string model [20] considers hyperon production as a re-combination of spectator quarks with participant quarks produced from the sea via the breaking of a color string. At string breaking a $\overline{q}q$ pair is produced with an orbital angular momentum which must be balanced by the spin of these quarks. The participant quark from the pair gives the hyperon a transverse momentum which depends on the orbital angular momentum of the pair, and also contributes to its spin, providing a correlation between the spin and $p_T$ of the final state particle. In this model the polarization is attributed to the participant quarks which are produced from the color field in the interaction; the polarization increases with the mass of the participant quark. Polarization of spectator quarks has also been incorporated into this model to account for the kaon beam data [21]. Another model [22], which we call the re-combination model, attributes the polarization to Thomas precession of a quark accelerated or decelerated in the color field. This model predicted the $\Lambda^0$ polarization in kaon beams.

The models consider a direct production of hyperons, while in inclusive experiments hyperons may be produced both directly and via decays of resonances. If this and other uncertainties are taken into account, both models provide a good agreement with polarization measurements in proton beams.

The dynamics of hyperon production in the forward direction by a hyperon beam are similar to those by proton beams. However the strangeness does not necessarily emerge in the production process but might originate from the strangeness already contained in the beam particle. Results from these measurements could give new insights into the systematics of polarization phenomena.

In this article we present measurements of the polarization of $\Lambda^0$, $\overline{\Lambda}^0$, $\Sigma^+$and $\Xi^-$ hyperons produced by a $\Sigma^-$ beam.

## 2 Experiment WA89

Experiment WA89 is performed with the charged hyperon beam of the CERN SPS. Its main purpose is to study the production of charmed baryons and to search for exotic states. The polarization studies were done in parallel with the main program.

Hyperons were produced by 450 GeV/c protons impinging on a 40 cm long beryllium target with a diameter of 0.2 cm. A magnetic channel consisting of 3 magnets with an integrated field of 8.4 Tm selected negative particles with a momentum of 330 GeV/c and a momentum spread of $\sigma(p)/p = 7\%$, produced at an angle to the proton beam smaller than 0.5 mrad. Therefore the beam particles were produced at $p_T < 0.2$ GeV/c and $\Sigma^-$ hyperons should have a polarization below 0.03. After a distance of 16 m the produced hyperons hit the experimental target, which consisted of a copper and a carbon block arranged side by side, each having a thickness corresponding to 3.7 % of an interaction length. The size of the beam at the target was 3.6 cm horizontally and 1.5 cm vertically, and its dispersion was 0.6 mrad in the horizontal plane and 1.0 mrad in the vertical plane. An average beam spill of 2.1 s contained about $1.8 \cdot 10^5$ $\Sigma^-$ hyperons and about $4.5 \cdot 10^5$ $\pi^-$ at the experimental target for an incoming intensity of $3.0 \cdot 10^{10}$ protons per spill. A transition radiation detector was used to discriminate online between $\pi^-$ and hyperons. The remaining high-momentum pion contamination was less than 10% of the hyperon sample, which contained mainly $\Sigma^-$ and about 2% of $\Xi^-$ hyperons.

Fig. 1 shows a sketch of the experimental setup used in the 1991 run of WA89 on which the data of this paper are based. The beam and the secondary particles were detected by 18 silicon micro-strip planes with 25 and 50$\mu$m pitch. Positioning the target about 14 m upstream of the center of the $\Omega$-spectrometer provided a 10 m long decay area for short living strange particles. The products of these decays along with the particles coming directly from the target were detected by 36 planes of drift chambers with a spatial resolution of about 300 $\mu$m. The particle momenta were measured by the $\Omega$-spectrometer [23] consisting of a super-conducting magnet with a field integral of 7.5 Tm and a tracking detector consisting of 30 MWPC planes inside the field area and 8 drift chamber planes at the exit of the magnet. The momentum resolution was $\sigma(p)/p^2 \approx 10^{-4}$ (GeV/c)$^{-1}$. Charged particles were identified using a ring imaging Cherenkov (RICH) detector [24]. It had a threshold of $\gamma = 42$ and provided $\pi/p$ separation up to about 150 GeV/c. Downstream of the RICH a lead glass electromagnetic calorimeter was positioned for photon and electron measurement [25]. In the polarization measurement it was used for $\pi^0$ reconstruction. The electromagnetic calorimeter was followed by a hadron calorimeter (not used in this measurement).

The trigger was relatively open. It selected about 20% of all $\Sigma^-$ induced interactions using multiplicities measured in scintillator hodoscopes and proportional chambers downstream of the magnet and correlations of hits in these detectors to select particles with high momenta. More than 2 particles at the exit of the magnet were required.

We present here results from the analysis of about 100 million events recorded in 1991.

## 3 Data analysis

In accordance with parity conservation in strong production processes, any non-vanishing polarization must be transverse to the production plane constructed from the beam and secondary particle momenta, thus the polarization axis is defined as: $\boldsymbol{n_{prod}} = \boldsymbol{k_{beam}} \times \boldsymbol{k_{part}}$ , where $\boldsymbol{k_{beam}}$ and $\boldsymbol{k_{part}}$ are the directions of the beam and the secondary particle. The distribution of daughter particles with respect to the polarization axis is parametrized by

$$\frac{dN}{d\Omega} \propto (1 + \alpha \mathcal{P}\ \boldsymbol{n_{prod}} \cdot \boldsymbol{\hat{k}_b}) \tag{1}$$

where $\mathcal{P}$ is the hyperon polarization, $\alpha$ is the asymmetry parameter of the weak decay and $\boldsymbol{\hat{k}_b}$ is the direction of the daughter baryon in the rest frame of the decaying hyperon.

The decays studied and the corresponding $\alpha$-parameters [26] are summarized in Table 1.

### 3.1 Event selection

Only hyperon decays upstream of the magnetic field of the $\Omega$-spectrometer were considered in this analysis, thus the minimum requirement for tracks of daughter particles was that they had reconstructed segments in the decay area drift chambers and in the $\Omega$-spectrometer. The interaction vertex was reconstructed with at least 3 charged tracks including the beam track. Candidates for hyperon decays were selected in the following ways.

1) In the search for $\Lambda^0 \to p\pi^-$ decays all combinations of positive and negative tracks were considered. Their point of closest approach had to be more than 50 cm downstream of the target. The momentum of the $\Lambda^0$ candidate had to be larger than 50 GeV/c. In order to reduce the background caused by $K^0_S \to \pi^+\pi^-$ decays, we excluded track combinations with a reconstructed $\pi^+\pi^-$ mass within $\pm 10$ MeV/c$^2$ of the $K^0$ mass.

2) The search for $\overline{\Lambda}^0 \to \overline{p}\pi^+$ decays was geometrically similar to the search for $\Lambda^0$ decays. However, the $\overline{p}$ candidates were rejected if they were positively identified as pions by the RICH detector, reducing the background by a factor of three. Since the average momentum of $\overline{\Lambda}^0$ is significantly lower than the momentum of $\Lambda^0$, they decay closer to the target. Therefore, the cut on the vertex position was 10 cm downstream of the target and no momentum cut was applied. The $K^0$ candidates were not excluded, however the contamination from $K^0_S \to \pi^+\pi^-$ decays was removed later by a background subtraction.

3) The search for $\Sigma^+ \to p\pi^0$ decay was more complicated [28]. The track of the $\Sigma^+$ was reconstructed in the micro-strip detectors while the track of the daughter proton was reconstructed in the decay area drift chambers and its momentum was measured in the spectrometer. The kink between them is small and the limited spatial resolution of the drift chambers did not allow the decay point to be reconstructed with sufficient accuracy to extract the signal from a large combinatoric background, using only these two tracks. Therefore, an iterative procedure using the proton and $\pi^0$ candidates was applied to search for the decay point. At first all $\pi^0$ candidates were reconstructed from two-photon combinations in the electromagnetic calorimeter assuming that $\pi^0$s were produced at the interaction vertex in the target. However a wide mass window was accepted for the $\pi^0$ candidates in order to include $\pi^0$ s from $\Sigma^+$ decaying downstream of the target. All positive tracks were accepted as proton candidates. For every combination of a $\pi^0$ – proton candidate the 3-momentum was calculated. This momentum vector, originating at the interaction vertex, defines a line which approximates the $\Sigma^+$ line-of-flight. Its point of closest approach to the proton track was taken as a new $\pi^0$ origin. This cycle was then repeated until the coordinates of the $\pi^0$ origin converged. This origin had to lie more than 20 cm downstream of the interaction vertex. The $\pi^0$ mass was checked and the $\Sigma^+$ candidate's trajectory was required to match a track in the micro-strip detector.

4) To look for $\Xi^- \to \Lambda^0\pi^-$ decays $\Lambda^0$ candidates within $3\sigma$ of the mass resolution from the proper particle mass [26] were accepted. In addition, the reconstructed decay point of a $\Xi^-$ candidate had to lie more than 50 cm downstream of the target, and its trajectory had to match a track in the micro-strip detector. Non-interacting $\Xi^-$ originating from the beam itself were observed as a peak in the momentum distribution of the $\Xi^-$ and were removed by the cuts on the interaction vertex.

The mass distributions of the selected decays are presented in Fig.2. The full statistics of the samples of hyperons along with the mass resolutions, mass selection cuts and the observed mean values of $x_F$ are given in Table 1.

### 3.2 Polarization analysis

For every decay studied the data were split into several $p_T$ intervals. The measured angular distributions are affected by the acceptance and the data analysis algorithm:

$$\frac{dN_{meas}}{d\cos\theta} = A(\cos\theta)(1 + \alpha\mathcal{P} \cdot \cos\theta) \tag{2}$$

where $\cos\theta = \boldsymbol{n_{prod}} \cdot \hat{\boldsymbol{k}}_{\boldsymbol{b}}$ and $A(\cos\theta)$ is the apparatus function, which also depends on other kinematic variables. To remove the apparatus functions a bias cancelling technique was applied, similar to one used in [27], exploiting the symmetry of the setup with respect to the horizontal plane (the magnetic field deflects the particles mainly horizontally). The distributions of eq.(2) were separately measured for pairs of azimuthal sectors of the hyperon direction, called "Up" and "Down". The coupled sectors were chosen such that they had similar acceptances for opposite values of $\cos\theta$:

$$U(\cos\theta) = \frac{dN_U}{d\cos\theta} = A_U(\cos\theta)(1 + \alpha\mathcal{P} \cdot \cos\theta) \tag{3}$$

$$D(\cos\theta) = \frac{dN_D}{d\cos\theta} = A_D(\cos\theta)(1 + \alpha\mathcal{P} \cdot \cos\theta) \tag{4}$$

where $A_U(\cos\theta) \approx A_D(-cos\theta)$ for $-1 < \cos\theta < 1$ and thus

$$\frac{\sqrt{U(\cos\theta)\cdot D(\cos\theta)} - \sqrt{U(-\cos\theta)\cdot D(-\cos\theta)}}{\sqrt{U(\cos\theta)\cdot D(\cos\theta)} + \sqrt{U(-\cos\theta)\cdot D(-\cos\theta)}} \approx \alpha\mathcal{P}\cdot\cos\theta \tag{5}$$

for $0 < \cos\theta < 1$. The statistical error of the measurement of the asymmetry $\alpha\mathcal{P}$ is about $\sqrt{3/N_{tot}}$, where $N_{tot}$ is the full number of events in both "Up" and "Down" distributions.

For the $\Xi^- \to \Lambda^0\pi^-$ decays the situation is more complicated since the daughter baryon $\Lambda^0$ is polarized. Assuming $CP$ conservation in the decay and no final-state interactions it follows [29]:

$$\boldsymbol{P}_{\boldsymbol{\Lambda}} = \frac{\gamma_\Xi \boldsymbol{P}_\Xi + (\alpha_\Xi + (1-\gamma_\Xi)\hat{\boldsymbol{\Lambda}} \cdot \boldsymbol{P}_\Xi)\hat{\boldsymbol{\Lambda}}}{1 + \alpha_\Xi \hat{\boldsymbol{\Lambda}} \cdot \boldsymbol{P}_\Xi} \tag{6}$$

where $\boldsymbol{P}_\Xi$ is the $\Xi^-$ polarization, $\gamma_\Xi^2 = 1 - \alpha_\Xi^2$ and $\hat{\boldsymbol{\Lambda}}$ is the direction of the $\Lambda^0$ in the $\Xi^-$ rest frame. This polarization causes an asymmetry of the $\Lambda^0$ decay:

$$\frac{dN_\Lambda}{d\Omega} \propto (1 + \alpha_\Lambda \boldsymbol{P}_{\boldsymbol{\Lambda}} \cdot \hat{\boldsymbol{k}}_{\boldsymbol{p}}) \tag{7}$$

where the $\hat{\boldsymbol{k}}_{\boldsymbol{p}}$ is the daughter proton momentum in the $\Lambda^0$ rest frame. Such an asymmetry combined with the detector acceptance for $\Lambda^0$ can add an additional distortion to the distributions of eq.(3) and (4) for $\Xi^-$, which is not compensated by the bias cancelling technique explained above. (If the $\Xi^-$ is not polarized the $\Lambda^0$ has a polarization only along its momentum $\boldsymbol{P}_{\boldsymbol{\Lambda}} = \alpha_{\Xi} \cdot \hat{\boldsymbol{\Lambda}}$ in the rest frame of the $\Xi^-$ , which does not cause such a distortion.) To correct for this distortion we used an iterative procedure, which started by ignoring a possible distortion. Using the bias cancelling technique a preliminary value of $\boldsymbol{P}_{\boldsymbol{\Xi}}$ was obtained. From this $\boldsymbol{P}_{\boldsymbol{\Lambda}}$ was calculated using eq.(6) and a weight was assigned to each event, inversely proportional to the decay probability given in eq.(7), making the $\Lambda^0$ decays effectively symmetric. The distortion caused by the $\Lambda^0$ polarization was measured to be lower than 0.03 for $\boldsymbol{P}_{\boldsymbol{\Xi}} = -0.20$ and the procedure converged rapidly. This procedure was checked on Monte Carlo generated data and convergence to preset values of $\boldsymbol{P}_{\boldsymbol{\Xi}}$ was obtained.

### 3.3 Systematic errors

The large aperture of the spectrometer provided a relatively flat apparatus function and, therefore, relatively little false asymmetries. False asymmetries, which were found and removed by the bias cancelling technique, were less than 5-10% for all decays considered.

Systematic errors were studied by: a) analyzing simulated data; b) studying the decay $K^0_S \to \pi^+\pi^-$, which is symmetric as a decay of a spinless particle; c) studying dependences of the measured asymmetries on azimuthal angles and decay paths.

The algorithm was checked on simulated data for $\Lambda^0$ and $\Xi^-$. $\Lambda^0$s were simulated with a polarization value of 0.05. The distribution of the kinematic variables ($x_F$ and $p_T$ ) of the particles was chosen to be similar to the measured ones. A set of 50 000 $\Lambda^0$ was produced, of those about 15 000 remained after the different cuts were applied for $K^0$ exclusion, vertex position and minimum $x_F$. Then the simulated data were analyzed in the same way as the real data. Finally a polarization value of $\mathcal{P} = 0.042 \pm 0.016$ was obtained in a good agreement with the input value. In the same way $\Xi^-$ production with a polarization of $\mathcal{P} = -0.20$ was simulated and a value of $\mathcal{P} = -0.22 \pm 0.03$ was reproduced.

The $K^0_S \to \pi^+\pi^-$ decay is relatively similar geometrically to the decays of $\Lambda^0$ and $\overline{\Lambda}^0$ and it provides a tool to check the algorithm of the analysis and possible false asymmetries. The $K^0$ analysis was performed in a similar way to the $\Lambda^0$ analysis. About 360 000 decays were selected. The measured decay asymmetry (Table 2) in the region $p_T >$ 0.6 GeV/c was consistent with zero: $0.002 \pm 0.007$. We obtain a limit of 1.2% on a false asymmetry for $K^0$ decays at 90% confidence level, which reflects the precision of the bias cancelling for our setup. Taking into account the $\alpha$-parameter of decays we estimate that the systematic error on the polarization of $\Lambda^0$ and $\overline{\Lambda}^0$ should not exceed 0.02.

We studied the variation of measured asymmetries with azimuthal angles and decay paths, and also with variations of applied cuts. No statistically significant variation was found.

### 3.4 Results

For all decays considered the method of bias cancelling was applied to the background corrected distributions and the results are presented in Fig. 3 and in Table 2. The $\alpha$-parameters used are given in Table 1 [26]. Only the statistical errors are shown.

The $\Lambda^0$ polarization averaged in the full observed range of $x_F$ is generally negative and the lowest measured value is $\mathcal{P} = -0.055 \pm 0.015$ at $p_T = 1.3$ GeV/c. Comparing this result with the proton beam measurements which demonstrated a nearly linear dependence of the polarization on $p_T$ and reach $\mathcal{P} \approx -0.2$ at $p_T \approx 1$ GeV/c one concludes that the sign of the polarization is the same but the absolute value is significantly lower. In proton beams the $\Lambda^0$ polarization grows with $x_F$. No significant $x_F$ dependence of the $\Lambda^0$ polarization is found in our experiment.

As in other inclusive experiments we could not distinguish for individual events between direct production of $\Lambda^0$ and production via decays of resonances. However, we were able to estimate the contribution from $\Sigma^0 \to \Lambda^0\gamma$ decays. For about 15% of the $\Lambda^0$ sample $\Sigma^0$ decays were observed using the electromagnetic calorimeter data. By simulation of the $\Sigma^0$ production we determined that $(23 \pm 2)\%$ of all selected $\Lambda^0$ came from $\Sigma^0$ decays. Thus the measured $\Lambda^0$ polarization with respect to its production plane is a superposition of the polarization of $\Lambda^0$ stemming from $\Sigma^0$ decays $\mathcal{P}_\Lambda^\Sigma$ and of the "true" polarization, $\mathcal{P}_\Lambda^{\text{true}}$: $\mathcal{P} \approx k \cdot \mathcal{P}_\Lambda^\Sigma + (1-k) \cdot \mathcal{P}_\Lambda^{\text{true}}$, where $0.15 < k < 0.25$ is the part of observed $\Lambda^0$s stemming from $\Sigma^0$ decays. Since $\mathcal{P}_\Lambda^\Sigma = -1/3 \cdot \mathcal{P}_\Sigma$ [30], where $\mathcal{P}_\Sigma$ is the polarization of $\Sigma^0$, we expect that $\mathcal{P}_\Lambda^\Sigma$ is relatively low. We have also measured $\mathcal{P}_\Lambda^\Sigma$ to be $0.022 \pm 0.008$ at $p_T$>0.6 GeV/c. Therefore the $\Sigma^0$ contribution to the overall $\Lambda^0$ sample attenuates the "true" $\Lambda^0$ polarization by a factor of about 0.75-0.85.

The $\overline{\Lambda}^0$ polarization is consistent with zero, similar to the proton beam measurements. No dependence of the polarization on the kinematic variables was found.

The $\Sigma^+$ polarization is low and generally negative, in sharp contrast to proton beam measurements ($\mathcal{P} \approx 0.15$ at $p_T \approx 1$ GeV/c). No significant dependence on $x_F$ was found.

The $\Xi^-$ polarization measured for the full observed range of $x_F$ is negative and reaches a value of $\mathcal{P} \approx -0.1$ at $p_T > 1$ GeV/c. A strong $x_F$ dependence of the polarization is found (Table 3 and Fig. 3e-f). At $x_F \approx 0.5$ a polarization of about $-0.2$ is observed, similar to the $\Lambda^0$ polarization in proton beams. In proton beams the $\Xi^-$ polarization grows slightly with the beam energy [5, 8], reaches values of about $-0.10$ at $p_T > 1$ GeV/c at 400 GeV, and does not show a significant $x_F$ dependence [8].

No statistically significant dependence of the measured polarizations on the target material (copper and carbon) was observed.

The selected beam sample contained about 10% of $\pi^-$ and 2% of $\Xi^-$ contamination. We expect the background from the $\pi^-$ contamination to be low for hyperons produced in the observed range of $x_F$ . Indeed, the ratio of differential cross-sections of $\Lambda^0$ production at $x_F = 0.2$ by proton and $\pi^-$ beams is about 5 [14, 31]. We assume that the hyperon production by hyperons in the forward hemisphere is enhanced even more than the production by protons. Thus the hyperon background from the $\pi^-$ content of the beam should be less than 2% and be negligible.

The $\Xi^-$ content of the beam should also be negligible for all studied hyperons except for $\Xi^-$. No measurement of a $\Xi^-$ production cross-section by a $\Sigma^-$ beam has been done, however a large inclusive cross-section of $\Xi^-$ produced by a $\Xi^-$ beam has been measured [32]. In order to estimate the ratio of $\Xi^-$ production cross-sections by $\Xi^-$ and $\Sigma^-$ beams we compared cross-sections of $\Sigma^-$ and $\Xi^-$ production by a proton beam [33]. Production of a hyperon with an additional $s$-quark at $x_F \approx 0.4$ is suppressed by a factor of about 10. Assuming that the $\Xi^-$ production by $\Sigma^-$ is suppressed by the same factor with respect to the production by $\Xi^-$ , we estimated the background from $\Xi^-$ produced by the $\Xi^-$ content of the beam to be about 15-30%, in the relevant range of $x_F$. The $s$-quark polarization is transferred to the $\Xi^-$ polarization with a dilution factor of 2/3. In $\Xi^-$ production by a $\Xi^-$ beam we expect $s$-quarks to be spectators. If these spectators are unpolarized, the $\Xi^-$ is unpolarized too. If they become polarized we expect their polarization to be similar to that of the $s$-quark in $\Lambda^0$ production by $K^-$ [16], namely positive. Thus, the effective polarization of $\Xi^-$ originating from $\Sigma^-$ may be stronger than the observed negative polarization of about $-0.20$ at $x_F \approx 0.5$.

### 3.5 Discussion

As discussed in the introduction the observed values of the hyperon polarization are strongly connected to the details of the production, mainly the origin of the strange quark. In our experiment the strangeness of the secondary hyperons may originate from the $\Sigma^-$ beam or be produced in the interaction. In the inclusive approach of this experiment there is no unambiguous way to distinguish between these processes, especially at low $x_F$ , since the associated strange particles may remain undetected. For the fragmentation region $x_F > 0.3$ we expect a considerable contribution of processes with at least one spectator quark transferred to the secondary hyperon, which may be the strange or a light quark.

For the $\Sigma^+$ production by $\Sigma^-$ only the $s$-quark can be the spectator quark. Thus, the $\Sigma^+$ polarization in a $\Sigma^-$ beam should have features, similar to the $\Lambda^0$ polarization in a $K^-$ beam. The $\Lambda^0$ polarization has been measured in a 176 GeV/c $K^-$ beam at $x_F \approx 0.5$ with the same inclusive approach [16], to be about $+0.50$. This polarization is attributed to the polarization of the spectator $s$-quark in the re-combination process. Taking into account that the $\Lambda^0$ spin practically coincides with the spin of the $s$-quark, while the $\Sigma^+$ spin is opposite to it with a probability of 2/3, one may expect a strong negative polarization of $\Sigma^+$ produced in a $\Sigma^-$ beam, however diluted with respect to the $\Lambda^0$ polarization by a factor of 1/3. The argument for a dilution holds also for $\Lambda^0$ and $\Sigma^+$ production in proton beams. The measured dilution factor in this case is about 0.5. Thus for our case an observable polarization of about $-0.25$ may be expected. Such a value is also predicted by the re-combination model [22], while the Lund model [20] predicts a low polarization. The low polarization of $\Sigma^+$ observed in our experiment does not show the strong polarization of the spectator $s$-quark, which appears to be seen in $K^-$ beam measurements.

For $\Lambda^0$ the re-combination model predicts the polarization to be about half of that observed in proton beams which would be about $-0.10$. The Lund model predicts a low polarization. We obtained a value significantly lower than one half of the proton beam results, if the full measured $p_T$ range is taken into account.

Both models considered predict the $\Xi^-$ polarization to be close to the polarization of $\Lambda^0$ produced in proton beams, in agreement with our results.

In summary, we have measured of the polarization of $\Lambda^0$ , $\overline{\Lambda}^0$ , $\Sigma^+$ and $\Xi^-$ inclusively produced in $\Sigma^-$ induced interactions at 330 GeV/c. This is the first measurement of polarization of hyperons produced by a hyperon beam. Only $\Xi^-$ show a significant polarization, comparable to the polarization of $\Lambda^0$ produced in proton beams. The Lund string model describes qualitatively all observations of this paper.

## Acknowledgements

We are indebted to J.Zimmer and the late Z.Kenesei for their help during all moments of detector construction and set-up. We are grateful to the staff of CERN's EBS group for providing an excellent hyperon beam channel, to the staff of CERN's Omega group for their help in running the $\Omega$-spectrometer and also to the staff of the SPS for providing good beam conditions. We thank G.Gustafson for a helpful discussion.

| # | Decay | $\alpha$ parameter | # events | $\sigma$ MeV/c$^2$ | mass cut MeV/c$^2$ | $<x_F>$ |
|---|---|---|---|---|---|---|
| 1 | $\Lambda^0 \to p\pi^-$ | 0.642 | 1800k | 2.7 | $\pm 6$ | 0.31 |
| 2 | $\overline{\Lambda}^0 \to \overline{p}\pi^+$ | $-0.642$ | 160k | 2.0 | $\pm 5$ | 0.11 |
| 3 | $\Sigma^+ \to p\pi^0$ | 0.98 | 45k | 6.0 | $\pm 10$ | 0.27 |
| 4 | $\Xi^- \to \Lambda^0\pi^-$ | $-0.456$ | 80k | 3.0 | $\pm 9$ | 0.31 |

Table 1: The analyzed decays

| $< p_T >$ GeV/c | Polarization | | | | asymmetry |
|---|---|---|---|---|---|
| | $\Lambda^0$ $< x_F >= 0.30$ | $\overline{\Lambda}^0$ $< x_F >= 0.11$ | $\Sigma^+$ $< x_F >= 0.27$ | $\Xi^-$ $< x_F >= 0.31$ | $K^0_S$ |
| 0.20 | $0.002 \pm 0.005$ | $0.013 \pm 0.023$ | $0.010 \pm 0.034$ | $-0.019 \pm 0.040$ | $0.002 \pm 0.011$ |
| 0.46 | $-0.004 \pm 0.004$ | $0.007 \pm 0.014$ | $-0.025 \pm 0.020$ | $0.001 \pm 0.027$ | $0.012 \pm 0.007$ |
| 0.73 | $-0.005 \pm 0.004$ | $0.008 \pm 0.016$ | $-0.045 \pm 0.022$ | $-0.055 \pm 0.030$ | $0.009 \pm 0.009$ |
| 1.03 | $-0.022 \pm 0.008$ | $-0.016 \pm 0.024$ | $-0.031 \pm 0.023$ | $-0.090 \pm 0.041$ | $-0.020 \pm 0.014$ |
| 1.32 | $-0.055 \pm 0.015$ | $0.040 \pm 0.041$ | $-0.051 \pm 0.035$ | $-0.092 \pm 0.064$ | $0.030 \pm 0.024$ |
| 1.80 | $-0.033 \pm 0.020$ | $0.014 \pm 0.068$ | | $-0.121 \pm 0.091$ | $-0.025 \pm 0.039$ |

Table 2: Polarization of hyperons and the asymmetry of $K^0_S$ measured for the full observed range of $x_F$. The mean values of $p_T$ in the selected $p_T$ intervals coincide for all particles. The errors indicated are statistical only.

| $< p_T >$ GeV/c | Polarization | | | $< x_F >$ | Polarization $< p_T >= 1.3$ GeV/c |
|---|---|---|---|---|---|
| | $< x_F >= 0.19$ | $< x_F >= 0.33$ | $< x_F >= 0.52$ | | |
| 0.20 | $0.058 \pm 0.057$ | $-0.065 \pm 0.080$ | $-0.106 \pm 0.079$ | 0.19 | $-0.003 \pm 0.055$ |
| 0.46 | $-0.002 \pm 0.038$ | $0.004 \pm 0.055$ | $0.011 \pm 0.054$ | 0.33 | $-0.133 \pm 0.069$ |
| 0.73 | $-0.011 \pm 0.042$ | $-0.129 \pm 0.062$ | $-0.071 \pm 0.061$ | 0.51 | $-0.280 \pm 0.069$ |
| 1.03 | $0.034 \pm 0.056$ | $-0.161 \pm 0.085$ | $-0.299 \pm 0.086$ | | |
| 1.32 | $-0.035 \pm 0.086$ | $0.064 \pm 0.143$ | $-0.322 \pm 0.141$ | | |
| 1.80 | $-0.113 \pm 0.121$ | $-0.366 \pm 0.201$ | $-0.063 \pm 0.214$ | | |

Table 3: Polarization of $\Xi^-$ measured for different $p_T$ and $x_F$ intervals. The last column contains the average values for the last three $p_T$ intervals. The errors indicated are statistical only.

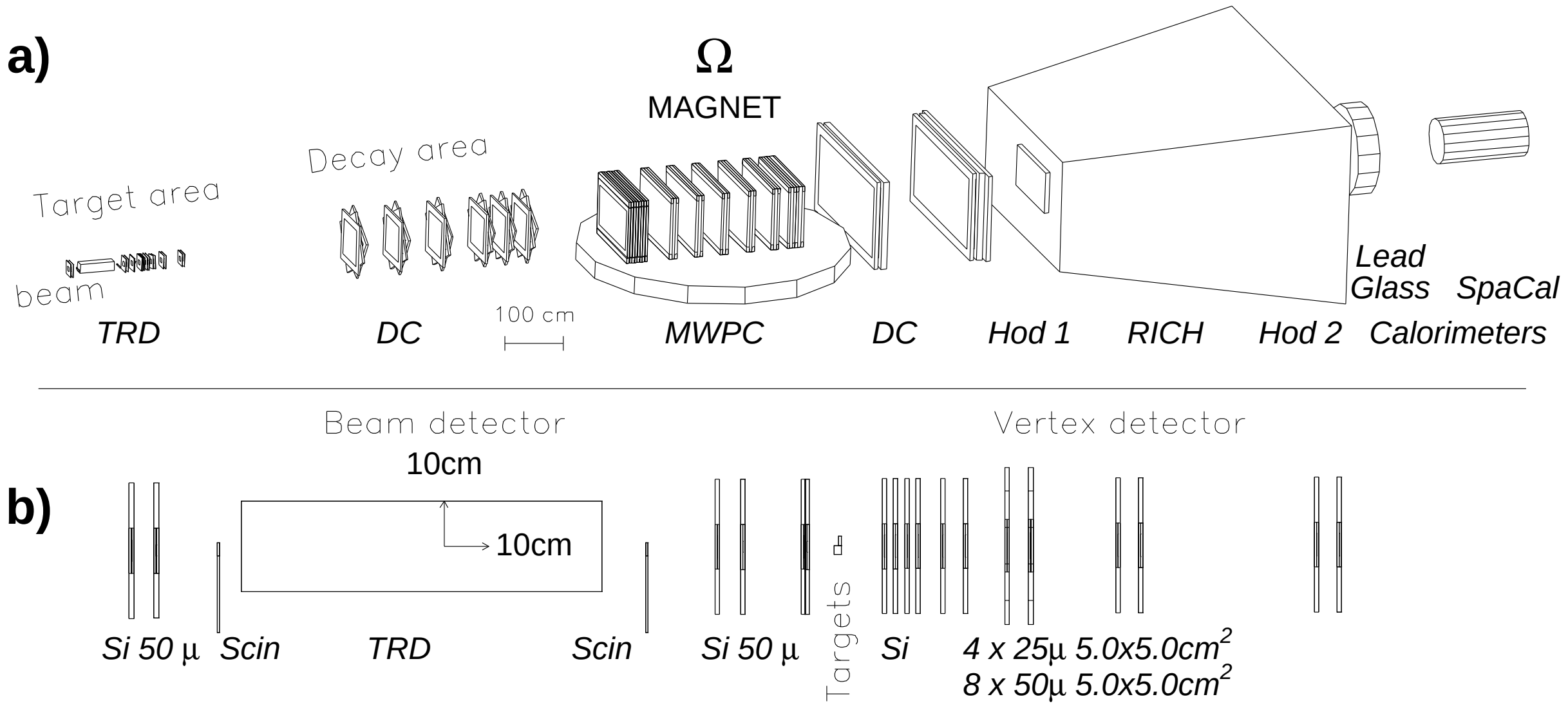

Figure 1: Setup of the WA89 experiment in the 1991 run, a) whole setup, b) the target area

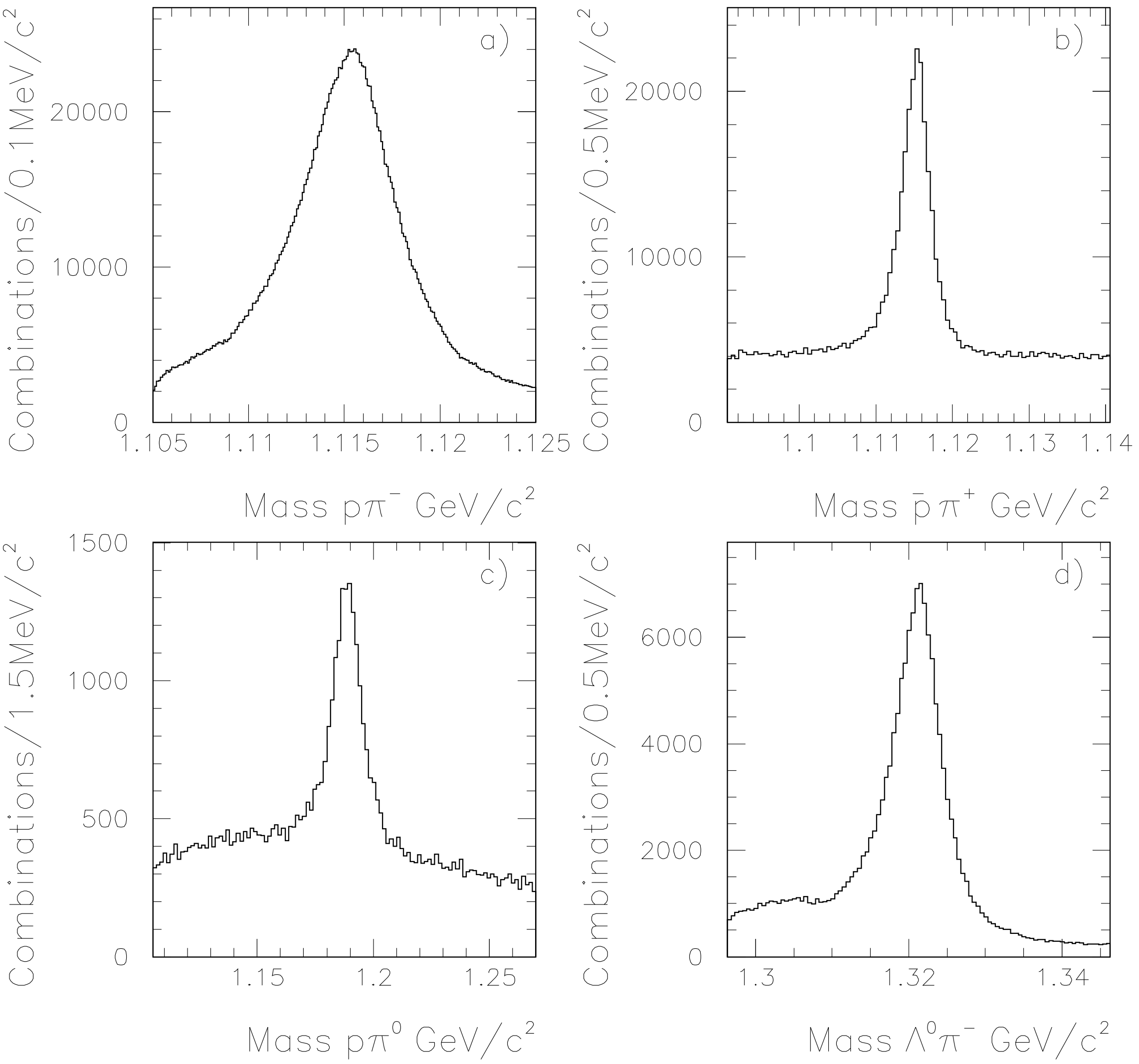

Figure 2: Invariant mass spectra of studied hyperons: a) $\Lambda^0$ , b) $\overline{\Lambda}^0$ , c) $\Sigma^+$ (presented for a part of the data sample used) and d) $\Xi^-$.

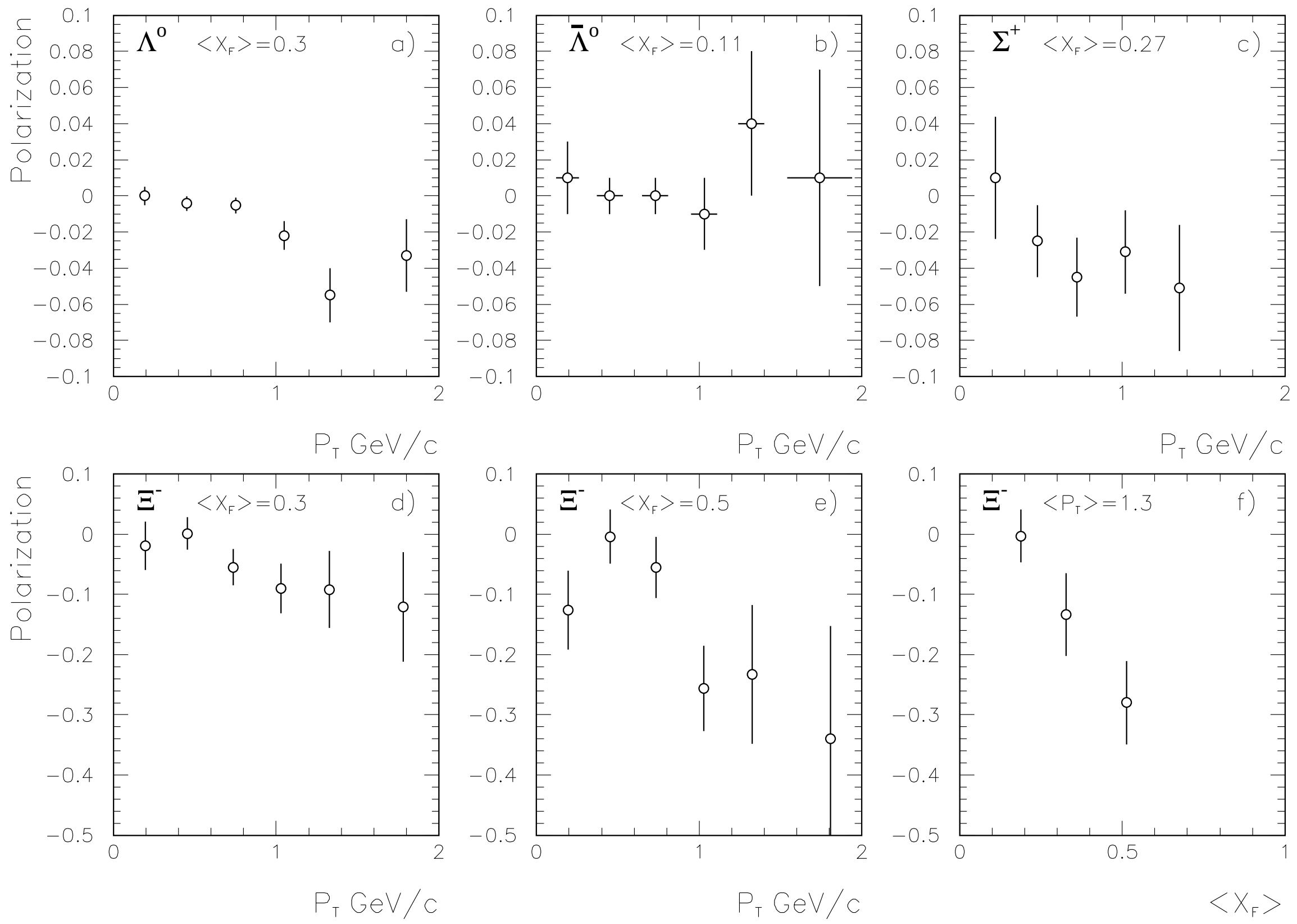

Figure 3: $p_T$ dependence of the polarization of a) $\Lambda^0$ , b) $\overline{\Lambda}^0$ , c) $\Sigma^+$ and d) $\Xi^-$ measured for the full observed ranges of $x_F$ (corresponding to mean values of $x_F$ of 0.30 , 0.11 , 0.27 and 0.31, resp.). e) polarization of $\Xi^-$ for $x_F > 0.3$, which corresponds to a mean value of $x_F$ of 0.5. f) $x_F$ dependence of the polarization of $\Xi^-$ for $p_T > 0.9$ which corresponds to a mean value of $p_T$ of 1.3. The errors indicated are statistical only.